\documentclass[sigconf, pbalance]{acmart}
\usepackage{fancyvrb}
\usepackage{caption}
\usepackage{subcaption}
\usepackage{tikz}
\usepackage{pgfplots}
\usepackage{microtype}
\usepackage{balance}
\usepackage{lipsum}
\usepackage{graphicx}
\usepackage{hyperref}
\usepackage{natbib}
\usepackage{listings}
\usepackage{xspace}
\usepackage{enumitem}
\usepackage{courier}
\usepackage{pgfplots, pgfplotstable}
\usepackage{multirow}
\usepackage[skip=0pt,belowskip=0pt,aboveskip=0pt]{caption}

\usetikzlibrary{automata,positioning}
\usepgfplotslibrary{statistics}
\usepgfplotslibrary{dateplot}
\pgfplotsset{compat=newest}
\copyrightyear{2026}
\acmYear{2026}
\setcopyright{rightsretained}  %
\acmConference[ICSE-SEIP'26]{48th International Conference on Software Engineering: Software Engineering in Practice}{April 12--18, 2026}{Rio de Janeiro, Brazil}
\acmBooktitle{48th International Conference on Software Engineering: Software Engineering in Practice (ICSE-SEIP '26)}
\acmDOI{}
\acmISBN{}

\begin{document}
\newcommand{\squeez}{}
\definecolor{DarkBlue}{RGB}{0,0,100}
\definecolor{LightBlue}{RGB}{0,100,200}
\definecolor{CornFlowerBlue}{RGB}{100,149,237}
\definecolor{Purple}{RGB}{102,51,153}
\newcommand*{\note}[3]{\textcolor{red}{[#1:} \textcolor{#2}{#3}\textcolor{red}{]}}

\newcommand*{\KM}[1]{\note{Ke}{DarkBlue}{#1}}
\newcommand*{\MM}[1]{\note{Matteo}{red}{#1}}
\newcommand*{\TK}[1]{\note{TK}{green}{#1}}
\newcommand*{\Cons}[1]{\note{Cons}{CornFlowerBlue}{#1}}
\newcommand*{\AH}[1]{\note{Akos}{LightBlue}{#1}}
\newcommand*{\DI}[1]{\note{Daniel}{Purple}{#1}}

\newcommand*\https[1]{\href{https://#1}{#1}}
\newcommand{\company}{WhatsApp}
\newcommand{\whatscode}{{\sc WhatsCode}\xspace}
\newcommand{\agentlessFixesNumber}{11}
\newcommand{\powershell}{PowerShell\xspace}

\title{WhatsCode: Large-Scale GenAI Deployment for Developer Efficiency at WhatsApp}

\newcommand*\affiliatinmeta[1]{\affiliation{%
  \institution{Meta}
  \city{London}
  \country{UK}
}}

\newcommand*\affiliatinmetampk[1]{\affiliation{%
  \institution{Meta}
  \city{Menlo Park}
  \country{USA}
}}

\newcommand*\affiliatinmetait[1]{\affiliation{%
  \institution{Meta}
  \city{Milan}
  \country{Italy}
}}

\newcommand*\affiliatinmetasw[1]{\affiliation{%
  \institution{Meta}
  \city{Stockholm}
  \country{Sweden}
}}

\author[K. Mao]{Ke Mao}
\affiliatinmeta{}
\email{kemao@meta.com}

\author[T. Kapus]{Timotej Kapus}
\affiliatinmeta{}
\email{kapust@meta.com}

\author[C. T {\AA}hs]{Cons T {\AA}hs}
\affiliatinmeta{}
\email{cons@meta.com}

\author[M. Marescotti]{Matteo Marescotti}
\affiliatinmeta{}
\email{mmatteo@meta.com}

\author[D. Ip]{Daniel Ip}
\affiliatinmeta{}
\email{danielip@meta.com}

\author[{\'A}. Hajdu]{{\'A}kos Hajdu}
\affiliatinmeta{}
\email{akoshajdu@meta.com}

\author[S. Cela]{Sopot Cela}
\affiliatinmeta{}
\email{scela@meta.com}

\author[A. Banerjee]{Aparup Banerjee}
\affiliatinmetampk{}
\email{abanerjee@meta.com}

\begin{abstract}

The deployment of AI-assisted development tools in compliance-relevant, large-scale industrial environments represents significant gaps in academic literature, despite growing industry adoption.
We report on the industrial deployment of WhatsCode, a domain-specific AI development system that supports WhatsApp (serving over 2 billion users) and processes millions of lines of code across multiple platforms. Over 25 months (2023-2025), WhatsCode evolved from targeted privacy automation to autonomous agentic workflows integrated with end-to-end feature development and DevOps processes.

WhatsCode achieved substantial quantifiable impact, improving automated privacy verification coverage 3.5× from 15\% to 53\%, identifying privacy requirements, and generating over 3{,}000 accepted code changes with acceptance rates ranging from 9\% to 100\% across different automation domains.  
The system committed 692 automated refactor/fix changes, 711 framework adoptions, 141 feature development assists and maintained 86\% precision in bug triage.
Our study identifies two stable human-AI collaboration patterns that emerged from production deployment: one-click rollout for high-confidence changes (60\% of cases) and commandeer-revise for complex decisions (40\%). We demonstrate that organizational factors, such as ownership models, adoption dynamics, and risk management, are as decisive as technical capabilities for enterprise-scale AI success. The findings provide evidence-based guidance for large-scale AI tool deployment in compliance-relevant environments, showing that effective human-AI collaboration, not full automation, drives sustainable business impact.
\end{abstract}



\maketitle

\section{Introduction}
The software engineering landscape is experiencing a paradigm shift with the integration of AI-powered development tools. While tools like GitHub Copilot~\cite{copilot2021}, Claude Code~\cite{anthropic2025claudecode}, Gemini CLI~\cite{google2025-gemini-cli}, and Cursor~\cite{cursor2025} have demonstrated strong potential for AI assistance in coding tasks, industrial deployment introduces unique challenges: managing large polyglot codebases across mobile and server platforms, accommodating diverse developer experience levels, meeting stringent privacy and compliance requirements, implementing reliability guardrails for global-scale systems, and addressing organizational adoption barriers. Despite growing industry interest, the practical deployment of comprehensive AI development systems in compliance-relevant, large scale cloud-based industrial environments represents significant gaps in academic literature.

This paper presents WhatsCode, a domain-specific AI development system deployed at WhatsApp, one of the world's largest messaging platforms with over 2 billion users~\cite{wa2025}. WhatsCode represents a comprehensive approach to AI-assisted development, extending beyond simple code completion to encompass privacy compliance automation, framework adoption, operational tasks, and quality assurance across mobile and server codebases, as well as new feature authoring. In particular, WhatsCode supports systematic implementation workflows that break down large features into multi-step tasks, coordinate cross-repo changes with guardrails (tests, static analysis, reviews) throughout execution.
Our work addresses a gap in understanding the organizational challenges and business impact of AI tool deployment in real enterprise environments with complex codebases, stringent compliance requirements, and diverse engineering teams. Unlike previous studies that focus on controlled environments or individual developer productivity, we report on 25 months of production deployment, providing empirical evidence of the organizational and business factors that determine deployment success.

\noindent This work makes the following key contributions to the field of enterprise AI-assisted software development:

\noindent \textbf{Empirical Enterprise Study:} We present a comprehensive 25-month longitudinal study of domain-specific AI development tool deployment at enterprise scale, encompassing both agentless and agentic approaches with quantitative business impact analysis across compliance-relevant domains.

\noindent  \textbf{Quantitative Impact:} We demonstrate substantial business value, including 3.5$\times$ improvement in privacy verification coverage (15\% to 53\%), identification of previously unknown requirements, generation of over 3{,}000 accepted code changes, with acceptance rates ranging from 9\% to 100\% across diverse automation domains.

\noindent  \textbf{Human-AI Collaboration Patterns:} We identify and characterize two stable collaboration modes that emerged from production deployment: \textit{one-click rollout} (60\% of cases) and \textit{commandeer-revise} (40\% of cases), providing empirical guidance for effective human-AI interaction design in enterprise environments.

\noindent  \textbf{Organizational Success Framework:} We discuss a three dimensional deployment framework encompassing technical architecture, organizational readiness, and risk management, demonstrating that responsibility attribution models, cultural adoption dynamics, and graduated autonomy strategies determine deployment success independent of raw algorithmic performance.

\noindent  \textbf{Compliance-relevant Deployment Practices:} We share empirically validated experience for AI deployment in regulated environments, including risk-stratified automation frameworks with four-level autonomy gradation and enhanced validation protocols addressing AI-specific failure modes identified through systematic incident analysis.

The remainder of this paper is organized as a chronological narrative of WhatsCode's evolution, driven by advancing GenAI capabilities and empirically discovered limitations: Section
\ref{sec:background} provides background and motivation; Section
\ref{sec:privacy_era} describes the initial Foundation Era (2023) with privacy automation and quantitative results; Section
\ref{sec:agentless_era} details the Agentless Expansion Era (2024) with deterministic workflows and business impact; Section
\ref{sec:agentic_era} presents the Agentic Evolution Era (2025) with workflows and results; Section~\ref{sec:challenges} discusses organizational and technical challenges across phases, and \ref{sec:synthesis} discusses answers to research questions through our empirical findings; Sections; \ref{sec:related} and \ref{sec:threats} present related work and threats to validity.
Finally, Section~\ref{sec:conclusion} concludes.

\section{Background and Motivation}
\label{sec:background}

\subsection{Enterprise AI Development Challenges}
Large-scale software organizations face unique challenges when adopting AI development tools. Unlike individual developers or small teams, enterprise environments require tools that can:

\begin{itemize}[leftmargin=*]
\item Operate over large, polyglot codebases and monorepo with multiple sub-repositories.
\item Respect domain constraints (regulatory and technical) by design.
\item Integrate with existing build, review, and release workflows.
\item Provide accountability and auditability of AI-produced changes.
\item Scale across teams with heterogeneous expertise and practices.
\item Mitigate reliability risks and blast radius effects in high-scale production environments.
\end{itemize}
WhatsApp's development environment exemplifies these challenges. The platform maintains separate codebases (in a mono-repo setup) for Android (Java/Kotlin), iOS (Swift/Objective-C), Web (JavaScript/Flow), and server components (Erlang/Hack/C++). Combined these have over a million lines of code. The engineering organization spans multiple teams working on features ranging from core messaging functionality to advanced AI features. All teams work under stringent privacy and security requirements due to the platform's global scale and nature of messaging data.

\subsection{Privacy Compliance as a Challenge}
Privacy compliance represents one of the most time-intensive aspects of software development at WhatsApp. The platform must comply with both external privacy regulations and internal privacy standards through a comprehensive risk review process. Each new project and significant code change undergoes privacy assessment, which requires developers to identify and document data collection and processing, confirm privacy requirements with verification methods, provide auditable evidence of policy compliance, and maintain up‑to‑date privacy impact assessments throughout the lifecycle.

Prior to WhatsCode deployment, privacy verification was largely manual, with only 15\% of privacy requirements covered by automated verification. This created significant developer overhead and bottlenecks in the development process, while also introducing risks of potential inconsistency across teams.

\subsection{Limitations of General-Purpose AI Tools}
While general-purpose AI coding assistants have shown promise, our preliminary evaluation revealed significant limitations for enterprise use. 
General models fundamentally lack understanding of patterns that characterize enterprise development. Examples include repository or language specific tooling, internal coding standards, frameworks and architectures. 
Without context engineering techniques like Retrieval-Augmented-Generation (RAG) and the integration with internal tools, they cannot access or reason about enterprise-specific documentation, policies, and historical decisions that inform development choices.

Moreover, these tools face substantial integration barriers with enterprise development workflows, review processes, and compliance systems. Quality variability becomes particularly problematic for complex enterprise patterns and regulatory requirements, while issues may emerge from potential exposure of proprietary code and inability to ensure generated code meets organizational standards.

These limitations motivated the development of WhatsCode as a domain-specific solution tailored to WhatsApp's unique requirements and enterprise constraints, designed to address the fundamental mismatch between general-purpose AI capabilities and enterprise development realities.

\subsection{Research Focus}
Based on systematic analysis of 25 months of WhatsCode deployment across WhatsApp's engineering organization, this study addresses two fundamental research questions:

\begin{description}[leftmargin=*,style=sameline]
\item[\textbf{RQ1:}] \textit{What quantifiable business impact can domain-specific AI development platforms achieve in compliance-relevant enterprise applications?}

\item[\textbf{RQ2:}] \textit{What organizational factors and human-AI collaboration patterns determine successful deployment of AI development tools in enterprise environments?}
\end{description}

RQ1 examines measurable business outcomes from AI tool deployment, focusing on productivity improvements, compliance automation effectiveness, code quality impacts, and organizational improvement in highly regulated enterprise contexts.
RQ2 investigates the non-technical determinants of AI tool adoption, including ownership attribution models, cultural adoption dynamics, risk management frameworks, and the emergence of stable human-AI collaboration patterns at organizational scale. 

These research questions emerged from our empirical observations, where organizational readiness and demonstrable business value proved to be the primary determinants of sustainable AI tool adoption, often outweighing technical capabilities in determining deployment success.

\section{Foundation Era (2023): Privacy Automation
}
\label{sec:privacy_era}
The foundation era of WhatsCode began in July 2023 as a Hackathon project addressing compliance challenges at WhatsApp. This section describes two complementary capabilities that established the foundational patterns for human-in-the-loop GenAI deployment in high-friction compliance workflows.

\subsection{Automated Requirement Mapping}
\label{sec:rq_mapping}

In our prior work, we presented PrivacyCAT~\cite{privacycat24}, a code analysis system to protect user privacy. A challenge was codifying privacy policies: mapping natural-language requirements to machine-verifiable checks. Prior to WhatsCode, this mapping was largely manual, with only 15\% of privacy requirements covered by automated verification.
The core problem was \textit{equivalent requirements}: different teams and stakeholders wrote semantically equivalent privacy requirements in different ways. For example, a requirement covered by our \textit{message-pair} dataflow property~\cite{privacycat24} could be phrased as "do not log a pair of numbers" or "do not log sender and receiver numbers." WhatsCode used LLMs to group such equivalent requirements and map them to existing verification methods.

\noindent \textbf{Technical Implementation:}
WhatsCode introduced a privacy-focused Retrieval-Augmented Generation (RAG) knowledge base to retrieve relevant internal context and reliably identify semantically similar requirements expressed in different natural-language forms. Mappings between privacy requirements and verification methods were stored in a JSON file checked into source control, initially seeded with pre-existing manual mappings for few-shot learning.

The system operated through daily automated scans of unmapped privacy requirements. Using the RAG system, WhatsCode proposed new mappings when confidence was high, directly editing the JSON file and creating change requests (diffs) for human review. Each diff included concise commit messages summarizing the proposed mappings, consolidating context that would otherwise require cross-referencing identifiers and numeric codes.

\noindent \textbf{Human-in-the-Loop Workflow:}
Human review typically followed two patterns: 1) \textit{one-click rollout} when proposals were correct; 2) \textit{commandeer-and-revise} when WhatsCode produced partially correct mappings that reviewers refined. 
The feedback were documented to a gatekeeper system for future reference and prevention.
This workflow integrated with existing source control, review and release processes, providing strong human oversight and building trust in WhatsCode outputs for a compliance-relevant domain.

\subsection{Privacy Review Discovery}
To further reduce privacy review effort and improve evidence reuse, WhatsCode introduced an assistant that discovers approved privacy reviews and attributes text or code to relevant precedents. Practitioners can invoke the tool via messaging or Web apps with as little as a one‑line query. The tool supports two interaction modes: 
\begin{itemize}[leftmargin=*] 
\item \textbf{Text-to-Review:} Developers describe a project or paste a review ID; the system retrieves the most relevant precedents and any available automated evidence.
\item \textbf{Code-to-Review:} Developers paste a code link to a file or line range. The system returns recommended approved reviews and reusable evidence through three attribution sources: 1) links between diffs and reviews; 2) File$\rightarrow$Project$\rightarrow$Feature$\rightarrow$Review linkages; and 3) a global semantic search over code content. 
\end{itemize}

\begin{figure}[tbp] \centering \includegraphics[width=\linewidth]{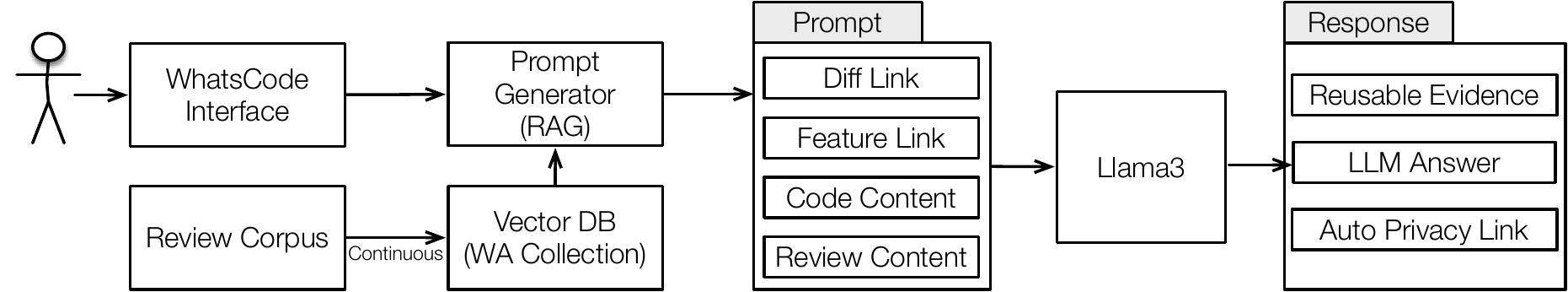} 
\caption{WhatsCode for privacy discovery.}
\squeez
\label{fig:lama-arch} 
\end{figure}

Figure ~\ref{fig:lama-arch} presents an overview of WhatsCode for privacy discovery. 
The system extends the solution as described in section~\ref{sec:rq_mapping}.
A continuously updated WhatsApp-specific review corpus is embedded into a vector database. A RAG \emph{prompt generator} constructs prompts that interleave: diff-level privacy review link, feature linking evidence, local code content, and relevant review content.
Prompts are sent to a Llama~\cite{touvron2023llama} model; responses are augmented with WhatsApp domain knowledge and returned to users as: 1) reusable evidence (links and artifacts for direct reuse), 2) an LLM answer summarizing applicability, and 3) an auto-generated link to the relevant privacy record when applicable.
The scope targets all approved WhatsApp privacy reviews, trading breadth for precision. 
By guiding reviewers to the right precedent and supporting artifacts, the command encourages reuse of evidence or reviews.

\subsection{Results: Contribution to RQ1}
\label{sec:foundation_rq1}

The Foundation Era provides the first empirical evidence for RQ1, demonstrating substantial quantifiable business impact in compliance-relevant applications:

\noindent \textbf{Privacy Automation Impact:}
WhatsCode's privacy requirement mapping automation launched in Q4 2023 with a phased rollout approach. After initial piloting, the system achieved significant quantitative improvements in reducing developer effort by turning many previously manual checks into automated verification:

\begin{itemize}[leftmargin=*]
\item Increased automated verification coverage from 15\% to 53\%,
\item Committed 290 diffs that added 1,535 new mappings,
\item of which 108 diffs without human edits (one‑click rollouts).
\end{itemize}

\noindent \textbf{Review Discovery Feedback:}
Because the review discovery capability was introduced as an assistive feature in chat and web surfaces, we evaluated it qualitatively rather than quantitatively. Feedback collected across Workplace threads indicated strong practitioner enthusiasm and perceived value for productivity (e.g., \textit{“I cannot emphasize enough how valuable and important it is.”}), with acknowledgements from privacy reviewers and engineers. The tool also showed usefulness beyond Engineering audiences: stakeholders suggested sharing with non‑Eng privacy partners for broader consumption. Through the tooling operations, we observed healthy human‑in‑the‑loop corrections: one user report flagged an over‑broad statement from the assistant; we refined answer templates to align with precedents and documented the change. These observations, while anecdotal, collectively suggest that Review Discovery reduces manual search time, encourages requirements and evidence reuse, and supports more efficient privacy decisions. 

\noindent \textbf{Organizational Impact:}
Beyond the quantitative and qualitative results, WhatsCode's foundation era established patterns: 1) the first deployment of GenAI within WhatsApp development process with demonstrable positive impact, establishing trust in AI's ability to address real problems; 2) a durable pattern for human-in-the-loop GenAI in high-friction workflows at Meta; and 3) catalyzing broader cultural adoption of AI assistance in developer workflows.

\subsection{Limitations and Evolution Drivers}
While effective for privacy, the RAG‑based approach alone was insufficient for broader developer efficiency goals. New use cases had to be curated and onboarded manually, and deeper integration with the end‑to‑end development toolchain was required to address general code quality, security, and operational needs. These limitations, alongside the demonstrated privacy efficiency improvements, motivated the next phase of WhatsCode: expanding beyond privacy into more systematic, multi‑domain automation integrated directly into developer tooling.

\section{Agentless Expansion Era (2024): Scaling Up
}
\label{sec:agentless_era}
\begin{table}[tbp]
    \caption{Number of closed diffs per lint type along with \% of closed out of AI generated diffs untouched by humans}
    \centering
    \footnotesize
    \renewcommand{\arraystretch}{1.1}
    \begin{tabular}{p{1cm}rrp{4.5cm}}
    \hline 
         \textbf{Type}      &  \textbf{Closed} & \textbf{Close\%} & \textbf{Description} \\ \hline
         Intents        & 708   & 31\% & Android \texttt{startActivity()} to internal wrapper with LLM-inferred metadata\\
         Broadcast      & 4     &  9\% & Similar to Intents, but on Android BroadcastReceivers\\
         Missing Thread & 17    & 58\% & Lint alerting at missing \href{https://developer.android.com/studio/write/annotations\#thread-annotations}{Android threading annotations}\\
         Wrong Thread   & 8     & 13\% & Similar to Missing Thread, but where the annotation is present, but wrong\\
         Plurals        & 8     & 29\% & Lint for \href{https://developer.android.com/guide/topics/resources/string-resource\#String}{String resources}, which expects annotations with the quantity. \\ %
         \powershell    & 252 & 83\% & Rewrite uses of unsafe \href{https://github.com/WhatsApp/power\_shell}{Erlang power\_shell} library\\
         Wrong Scope    & 6     & 11\% & Shows the uses of incorrect \href{https://kotlinlang.org/docs/scope-functions.html\#}{Kotlin scope functions}\\
         ClangTidy      & 65    & 18\% & Clang-Tidy\href{https://clang.llvm.org/extra/clang-tidy/}~lints on C/C++/ObjC codebase, mostly for inconsistent nullability annotations\\
         SwiftLint      & 15    & 100\% & \href{https://github.com/realm/SwiftLint}{SwiftLint} lints on Swift codebase related to performance optimization\\
         Infer          & 90    & 87\% & \href{https://fbinfer.com/}{Infer} issues on ObjC codebase related to memory leaks, nullability and perf\\
         Convert        & 105    & 41\% & Usage of internal Erlang library for converting data types (ie. string to integer)\\ \hline
    \end{tabular}
    \label{tab:agentless-lint-results}
    \vspace{-4mm}
\end{table}

Building on privacy automation success, 2024 marked WhatsCode's expansion into developer efficiency through deterministic, "pre-built" workflows where some steps would involve calling an LLM. 
This expansion implemented developer-designed workflows where LLMs operated within tightly constrained parameters. 
We adopted \textit{agentless}~\cite{agentless} as the term to describe this work, but we were unaware of their contribution at the time of undertaking this work.

\subsection{Agentless for Fixing Lint Issues}
We designed and implemented an agentless autofix system for lint issues for WhatsApp codebases. The high level goal of the system is to reduce the amount of lint issues reported on our code bases. The system would crawl existing issues, attempt to fix them and if successful create commits for humans to review and check into the code base.
This system followed a deterministic, prescribed workflow designed by programmers rather than allowing the LLM to make dynamic decisions about execution flow. In other words the agentless system implements a static directed graph of operations where the LLM's role is limited to code transformation within a predefined workflow:

\noindent  \textbf{Prompt Creation:} The system crafts LLM prompts using three components: 1) code snippet(s) extracted from 10-20 lines before and after all code locations associated with a lint issue; 2) lint descriptions matching developer-facing error messages; and 3) predefined templates based on lint type with expanded explanations. Previous validation attempts are included when available.

\noindent \textbf{GenAI Processing:} LLM generates modified code snippet and optional deterministic actions. The deterministic actions enable predetermines changes outside of the code snippet. The examples we implemented were adding imports at the top of the file and adding a dependency to the build file. 

\noindent \textbf{Code Integration:} Modified snippets are re-inserted using a simple heuristic matching the first and last two lines to locate insertion points. Any LLM-requested deterministic actions are applied.

\noindent \textbf{Validation Loop:} The system executes linters, compilation, and limited test suites. Failed validation triggers error feedback into the prompt, looping up to five iterations. Success advances to diff generation.

\noindent \textbf{Diff Generation:} Create and submit diff for review.

\subsection{Results: Contribution to RQ1}
The Agentless Era extends the quantifiable business impact evidence for RQ1, while revealing important boundaries for deterministic AI workflows. We deployed this system on \agentlessFixesNumber\ lint types. Their descriptions and results are summarized in Table~\ref{tab:agentless-lint-results}.
The lints were chosen based on business needs as well as feasibility (especially LLM capabilities) at the time. 

Diffs are {\em closed} when they are committed to master. The close\% was determined by dividing the number of diffs that appear in the history of master branch with all relevant diffs that were published by the agent. 

These results show the agentless approach can be deployed on a wide variety of issues spanning multiple languages (Java, Erlang, C/C++/Objective-C, Swift, and Kotlin). It achieved real impact, landing over 1000 code changes across \agentlessFixesNumber\ different lint types.

Three lints types stand out from Table~\ref{tab:agentless-lint-results}: \powershell, Intents and Infer. These three have a high number of closed diffs and high close rate when compared to the others. For these three lint types we worked with lint issue owners to eliminate these three lints, whereas we deployed  other lint types out of our own initiative. Our experience shows that while it is possible to deploy these approached in an ad-hoc way, they can only achieve large scale impact if there is sufficient will to eliminate those lints.

When comparing \powershell and Infer versus Intents lint we also note a large difference in close rate between them. This is because we took different approach for assigning reviewers to those diffs. \powershell and Infer lints had a dedicated team of reviewers whereas the Intents fixes were reviewed by code owners throughout the organisation.  For Intents lints this meant diffs were ignored more, which would eventually lead them to be abandoned and recreated, which decreases the close rate.

For the \powershell case there was a stated goal of removing all problematic instances.
Starting out there were over 3000 direct instances that needed to be rewritten. For consistency in Table~\ref{tab:agentless-lint-results}, the number of \powershell diffs generated by AI and untouched by humans is 304. In total, 632 \powershell diffs were generated. 
\begin{itemize}[leftmargin=*]
\item 252 were accepted with no modifications.
\item 312 were commandeered and modified by a human before acceptance.  The changes were, in general, minor, e.g., to fix formatting or naming. Moving towards the goal was more important than a high coverage of automated diffs.  These cases resulted in changes to the prompt and/or supporting code, improving future diffs.
\item 68 were abandoned; 16 duplicates, 52 AI code beyond saving.
\end{itemize}
The success rate, including the human assisted diffs, was thus $(252 + 312)/(252+312+68)=89\%$. This did not cover all the instances that needed to be corrected.  The tail comprised changes that were deemed too rare or too difficult to be done by AI (with tools at the time) or simply done manually for the sake of closing the task.

\subsection{Limitations and Insights}

The agentless era provided empirical evidence about the boundaries of deterministic AI workflows in enterprise environments. It showed good performance for problems with clear input/output boundaries and established solution patterns. It demonstrated ability to process large volumes of similar problems with small amount of engineering time investment.

The main difficulty we encountered with the agentless systems was scaling it to more problems that our engineers asked us to solve. This was due to agentless solution-space being limited to single-file, single-context problems. Adapting to a new problem type required significant manual template development, which required in-depth knowledge of the system and thus was not widely adopted outside of our team. 
The limited capacity available at the time in combination with the goal of moving forward lead to compromises in the form of commandeering diffs and making manual changes rather than rerunning (possibly after changing the prompt or support code).

Our empirical evaluation revealed scalability constraints: the highly variable success rates across lint types, coupled with substantial prompt engineering overhead for each new use case, demonstrated that agentless architectures fundamentally could not satisfy WhatsApp's comprehensive developer efficiency requirements.
These empirically validated limitations, combined with concurrent advances in AI capabilities and increasing organizational confidence in AI-assisted development workflows, established the theoretical and practical foundation for architectural evolution toward sophisticated agentic systems capable of multi-step reasoning and adaptive task execution.

A comparison of AI based analysis vs traditional program analysis is not a subject of this paper. An astute reader, however, might observe that some of the lint types in Table~\ref{tab:agentless-lint-results} could be candidates for traditional approach. For some lint, e.g., \textit{Convert} and \textit{Wrong/Missing Thread}, we estimate a traditional approach could work with some effort. We observe that fixing these with traditional approaches was not undertaken, but speculate that using AI assisted analysis has a significantly lower barrier to entry than traditional analysis.

\section{Agentic Evolution Era (2025): Autonomous %
}
\label{sec:agentic_era}

\begin{figure*}[t]
  \centering
  \includegraphics[width=15cm]{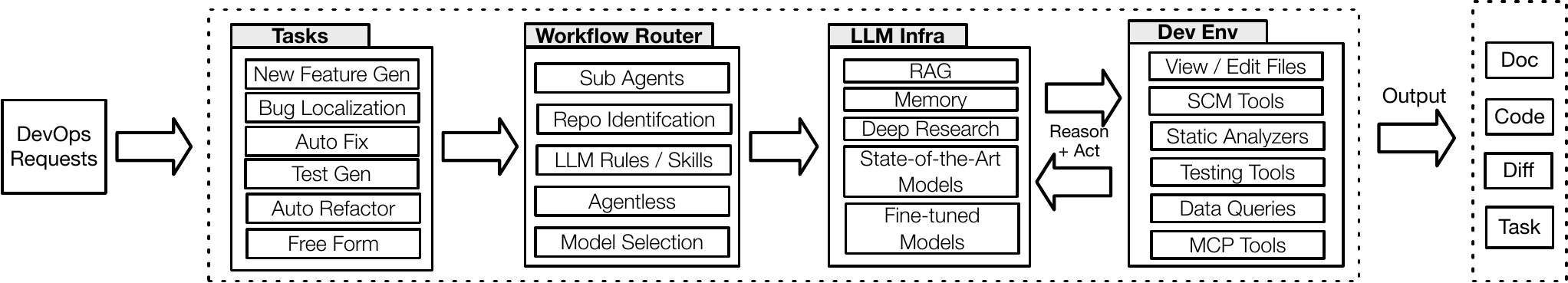}
  \caption{Overview of WhatsCode Agentic Workflow}
  \label{fig:agent}
  \squeez
\end{figure*}

WhatsCode's evolution in 2025 introduced sophisticated agentic systems capable of autonomous task orchestration, dynamic context assembly, and intelligent tool usage. This era represents a fundamental paradigm shift from reactive AI assistance toward proactive intelligent automation, addressing complex multi-step workflows that characterize enterprise software development at scale. Unlike previous approaches that required explicit human guidance for each step, the agentic system demonstrates emergent problem-solving capabilities through planning-driven execution and adaptive context management.

\subsection{Agentic Architecture}
Addressing fundamental limitations observed in agentless approaches, WhatsCode's agentic architecture employs a layered design separating concerns across distinct subsystems. The system transforms heterogeneous developer intents into robust, auditable, and policy-compliant workflows through four cooperating layers illustrated in Figure~\ref{fig:agent}.

\noindent \textbf{Task Layer:}
Incoming requests are normalized into canonical task types such as New Feature Authoring, Auto Fix, Test Generation, and Free-Form tasks, providing consistent interfaces for diverse developer needs.

\noindent \textbf{Workflow Router:}
The router converts task intents into executable plans and selects appropriate execution components. \textit{Repo-specific Coding Workflows:} Repository-aware flows injecting codebase rules, owners, and build/test conventions.
\textit{Sub Agents:} Multi-step planners delegate to specialized sub agents e.g., summarization, with memory usage and tool selection for open-ended work.
\textit{Agentless Workflows:} Sub agent could employ agentless workflows with deterministic, pattern-driven pipelines for high-precision changes (e.g., lint fixes and codemods).
\textit{Model Selection:} Chooses among base models of third-party vendors  based on task and sub agent types, latency/quality requirements, and data-governance constraints. Although feasible for autonomous selection, we rely on human configurations for transparency and simplicity. 

\noindent \textbf{LLM Infra:}
LLM infrastructure provides foundations supporting agentic workflows: \textit{RAG} enabling dynamic retrieval from code repositories and technical documentation, \textit{Deep research} capabilities for comprehensive information synthesis across heterogeneous knowledge sources, and \textit{Memory} management that maintain context across extended multi-step workflows with automatic compaction when approaching token limits. WhatsCode leverages various LLM models provided by Meta AI infra.

\noindent \textbf{Developer Environment:}
The agent operates through a unified tool interface abstracting common developer operations such as file editing and source control. Tool calls return results feeding back into plan state and prompting the next reasoning step through an iterative loop where the model reasons over current context, invokes tools, interprets results, updates plans and produces artifacts.

\subsubsection{Multi-Step Task Execution Capabilities}
The agentic system demonstrates sophisticated autonomous capabilities through a planner-executor architecture that decomposes complex enterprise development requests into structured, executable workflows. This approach addresses a fundamental limitation of previous systems: the inability to maintain coherence across extended task sequences while adapting to intermediate outcomes.
The system employs explicit task templates that scaffold complex workflows into ordered steps, each with descriptions and actions including tools that can be used. Key autonomous execution patterns include:

\noindent \textbf{Repo-wide Refactoring:} Large-scale refactoring operations demonstrate the system's ability to reason about dependencies and coordinate changes across module boundaries. The workflow encompasses code analysis per codebase, strategic planning that considers test dependencies, coordinated execution of changes with validation through automated testing, static analysis, reviewer and task access to support broader adoption.

\noindent \textbf{Review-driven Iteration:} For WhatsCode authored diffs, a particularly capability involves autonomous response to code review feedback. When human reviewers request modifications to agent-generated code, the system can autonomously interpret review comments, implement requested changes, and resubmit updated implementations for reducing the overhead of human-AI collaboration cycles.
In practice we also encourage reviewers to commandeer AI authored diffs for efficient iterations and keep human in the loop for recognition and accountability. 

\subsubsection{Dynamic Context Assembly}
A component of the agentic approach lies in its dynamic context assembly mechanism, which adaptively gathers and synthesizes relevant information based on task requirements. This represents a significant advance over agentless systems that rely on static, pre-specified context.
The system employs three complementary strategies for context management:

\noindent \textbf{Multi-Source Retrieval:} A hybrid of RAG and deep research capabilities synthesize information across heterogeneous internal knowledge sources, including codebases, Wikis, technical documentation and posts. This retrieval process operates on-demand, allowing the system to gather context precisely when needed rather than loading all potentially relevant information upfront.

\noindent \textbf{Dynamic Rule Injection:} The system discovers and applies domain-specific guidelines through repo/project rules that are dynamically loaded based on file access, rule scope, and user prompts. These rules enable internal technical decisions, workflows, coding standards, and domain constraints that are automatically injected into the reasoning process when relevant contexts are detected.

\noindent \textbf{Memory Management:} To support long-running, multi-step workflows, the system maintains persistent memory across interaction sessions with automatic context compaction when approaching token limits. This enables resumption of complex tasks while preserving progress and decision history, addressing the context window limitations.

\subsubsection{Enterprise Tool Orchestration}
The agentic system's tool orchestration capabilities represent a significant advancement in enterprise development automation, enabling seamless integration with existing development infrastructure while maintaining auditability.
The orchestration strategy employs three key principles:

\noindent \textbf{Policy-Gated Operations:} Tool interactions are mediated through policy enforcement mechanisms that validate operations before execution and provide guardrail feedback to the reasoning system. This approach ensures that automated actions comply with organizational constraints while enabling the system to learn from policy violations. One example is file search in large mono repo could be very computationally intensive and we have gated global searches.

\noindent \textbf{Allowlisted Command Execution:} The system operates within a constrained action space defined by carefully curated command allowlists that are tailored per repository and development context. This constraint mechanism balances capability with safety, preventing potentially harmful operations while enabling comprehensive automation of approved development tasks.

\noindent \textbf{Provenance and Auditability:} All automated actions generate detailed provenance tracking and create reviewable artifacts, enabling human oversight and debugging. This transparency mechanism is important for enterprise adoption, as it maintains accountability while reducing the need for constant human supervision.

\subsubsection{End-to-End Feature Development}
\label{sec:feature_development}

\begin{figure}[tbp]
  \centering
  \includegraphics[width=8cm]{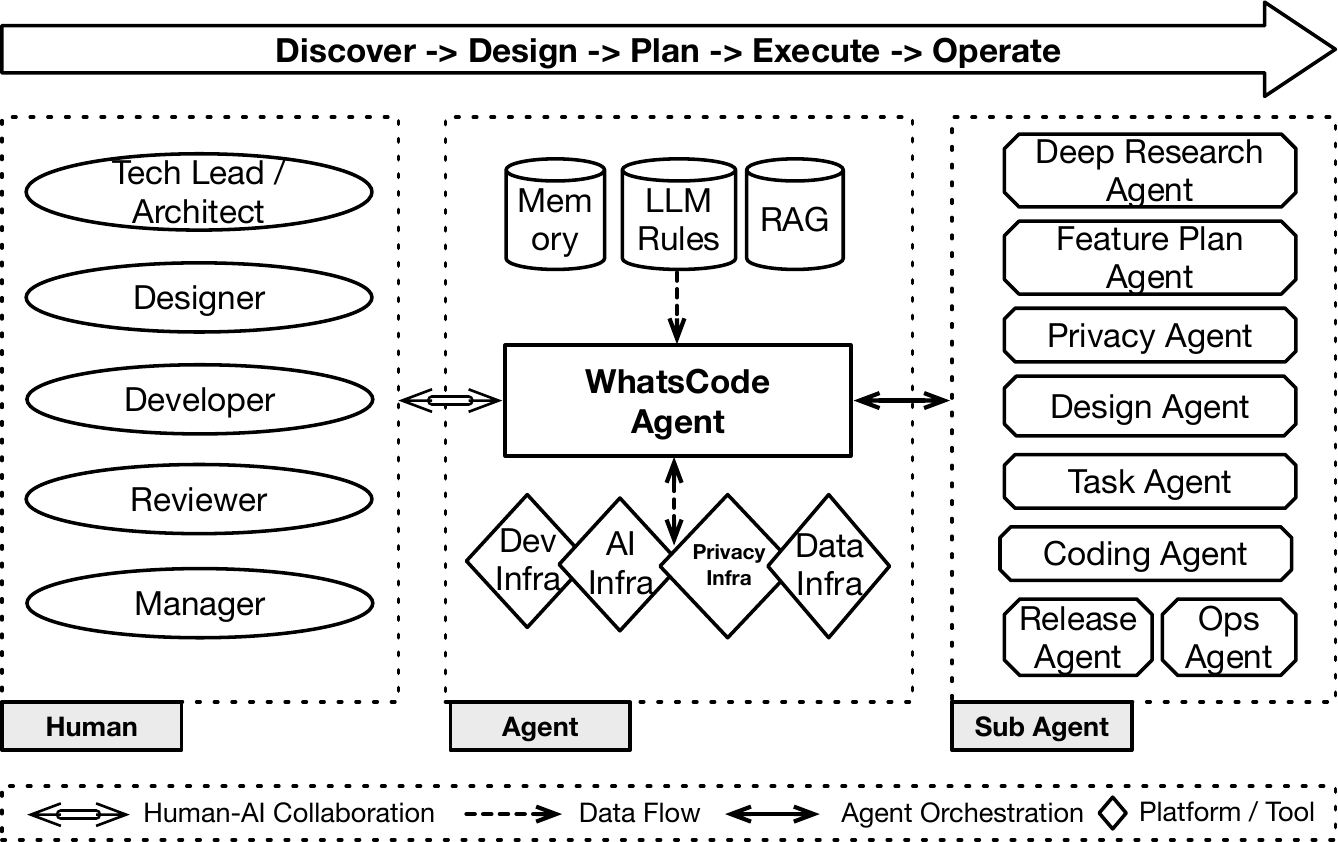}
  \caption{WhatsCode for end-to-end feature development.}
  \squeez
  \label{fig:feature}
  \vspace{-4mm}
\end{figure}

WhatsCode agent provides a sophisticated capability in orchestrating complete feature development lifecycles through a structured five-phase workflow encompassing discovery, design, planning, execution and operation, as illustrated in Figure~\ref{fig:feature}. 
This systematic approach addresses the complexity of enterprise feature development, where individual changes often require coordinated modifications across multiple repositories, adherence to organizational policies, and integration with existing development infrastructure. The system demonstrates early ability to decompose small to medium sized features into manageable multi-step tasks, maintaining implementation consistency across development cycles spanning multiple weeks. Through progressive complexity scaling, WhatsCode handles features requiring multi-stakeholder coordinate while preserving architectural coherence and provides support to ensure compliance with established engineering standards. 
This capability is supported by the cross-session memory, LLM rules and RAG which maintains context continuity across extended development periods, and its integration with task orchestration frameworks that coordinate multi-agent workflows based on task complexity and domain expertise. The main WhatsCode agent may delegate to sub agents which may be built on top of existing WhatsCode or other Meta agents.

\subsubsection{AI for Operations}
Engineers requested further integration of AI to support non-coding tasks. In response, we developed a bug triage and post-deployment incident investigation assistant. The assistant is initially agentless and has evolved into part of agentic workflow as demonstrated in Figure~\ref{fig:agent}.

\noindent \textbf{Bug triaging:}
The bug triaging bot is capable of reading bug reports and assigning task priorities according to the same guidelines used by our triage specialists. Although different service-level agreements (SLAs) exist for varying bug severities, engineers often struggled to meet the short SLAs associated with high-priority bug reports. Manual triaging could take one to two days, delaying assignment to the appropriate owner and reducing the time available for actual bug resolution. The introduction of the bot significantly improved the process by promptly routing high-priority issues to the correct owner, thereby enhancing SLA compliance and allowing engineers to focus more on fixing bugs.

The system reached 90\% precision for triaging high-priority bug after going through a few difficulties:
\begin{itemize}[leftmargin=*]
\item The system struggled with instructions that require hard-to-infer knowledge, such as reproducibility. Minor tuning of the bug report flow improved the triaging precision by a good margin. %
\item System tuning, including prompt changes and new context, should always be benchmarked. It is tempting to iterate rapidly to address specific cases that perform poorly, improvements in particular cases come at the cost of overall performance.
\end{itemize}

\noindent \textbf{Post-Deployment Incident Investigation:} Our incident investigation system successfully identified root causes for 21\% of production outages across diverse codebases, demonstrating the potential for AI-assisted debugging in complex distributed systems.

\noindent \textbf{Expanding Agent for oncall:} Oncall tasks and alerts tend to be more ambiguous, making them challenging for agentless AI systems, which are less likely to perform effectively in such contexts. To address this, we built upon an agent framework that already possesses substantial organization-specific context. However, the tools and integrations required for effective oncall investigation differ significantly from those needed for code modifications. Integrating with a wide range of systems—such as telemetry, logging platforms, and comprehensive runbooks—substantially enhances the performance of oncall support.

\subsection{Results: Contribution to RQ1}
The agentic evolution achieved unprecedented scale and scope, demonstrating measurable improvements in development efficiency, code quality, and operational automation across enterprise software development workflows.

\begin{figure}[tbp]
\begin{center}
\includegraphics{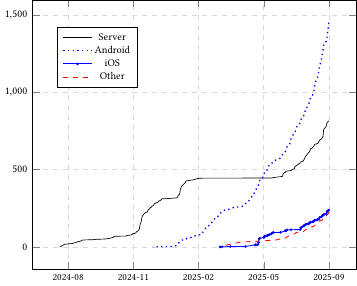}
\caption{WhatsCode committed diffs by platform over time.}
\squeez
\label{fig:multi_series_diffs}
\end{center}
\end{figure}

As shown in Figure~\ref{fig:multi_series_diffs}, since 2024 agentless era, committed diffs accelerated sharply after the agentic rollout (Q2 2025), with Android exhibiting the fastest growth, rising from near zero to approximately 1{,}500 diffs by September 2025. Server changes increased steadily with punctuated jumps due to organic adoption for more complex tasks like feature authoring and test generation, reaching approximately 800 diffs, while iOS and “Other” (e.g., configuration) platforms showed later but consistent adoption. The series reveal a breakpoint around May followed an organization-wide rollout. 
We attribute the trajectory change in Q3 2025 to incremental releases of WhatsCode support for developing new features with agentic capabilities like deep research and LLM rule improvements.
The overall time series indicates compound growth in code contributions and sustained usages, aligning with our agentic design’s emphasis on repo-aware and self-onboarding workflows for organic growth.

\subsubsection{Comprehensive Development Automation}
Our deployment results also demonstrate the system's effectiveness across multiple dimensions of enterprise development. These results demonstrate that agentic architectures can achieve quantifiable business impact by addressing complex, multi-step enterprise workflows that were previously impractical for AI automation.

\noindent \textbf{Adoption and Scale:} The system achieved significant organic adoption with approximately 20\% weekly active usage among WhatsApp engineers across Commandline and Web-based interfaces, and integrated development environments like review tools or work chat apps. This adoption rate represents substantial engagement for an enterprise AI system, particularly given the high technical sophistication of the user base.

\noindent \textbf{Code Quality Impact:} Quantitative analysis reveals substantial improvements in code quality metrics, most notably demonstrated by one Media team's code quality score (measured by detected lint issues) improvement from 45\% to 64\%. This 42\% relative improvement occurred over a sustained period while the system was actively assisting with fixing tasks, suggesting that agentic automation can contribute positively to long-term codebase health.

\noindent \textbf{Large-Scale Refactoring:} The project demonstrated capability for complex changes across multiple repositories, including the removal of 100s of obsolete A/B experiment and execution of extensive backend API migrations. These operations require understanding of cross-component dependencies and careful orchestration of changes to maintain system stability, representing tasks that traditionally require significant human effort and coordination.

\subsubsection{Comparative Evolution Analysis}

\begin{table}[t]
\captionsetup{width=.75\textwidth}
\caption{WhatsCode Evolution: Impact Analysis Across Eras}
\begin{center}
\scriptsize
\renewcommand{\arraystretch}{1.2}
\begin{tabular}{l|r|r|r}
\hline
\textbf{Metric} &
\textbf{Foundation} &
\textbf{Agentless} &
\textbf{Agentic (Era)} \\
\hline Code Changes Generated & 290 & 900+ & 2,000+ \\
Domain Coverage & 1 (compliance) & \agentlessFixesNumber~(dev efficiency) & 15+ (enterprise-wide) \\
Task Complexity Scope & Single-domain & Constrained patterns & Multi-step workflows \\
Context Assembly & Static & Pre-specified & Dynamic \\
Human Oversight Model & Collaborative & Supervisory & Graduated autonomy \\
Adaptation Capability & Rule-based & Template-based & Learning-enabled \\
Planning Capability & Human & Implicit & Explicit multi-step \\
Cross-Session Continuity & None & None & Memory-enabled \\
\hline
\end{tabular}
\label{tab:evolution}
\end{center}
\vspace{-6mm}
\end{table}

Table~\ref{tab:evolution} presents comparative analysis across WhatsCode's architectural phases, demonstrating empirically-driven progression where each phase addressed specific limitations while preserving successful capabilities. The agentic era's improved adaptability and planning capabilities enable significantly more complex tasks, though at increased computational overhead and need for oversight mechanisms.

The evolution preserves specialized approaches for tasks where simpler methods remain effective, suggesting enterprise AI deployment benefits from hybrid architectures rather than monolithic solutions. Table~\ref{tab:metrics} summarizes quantitative metrics from 25 months of WhatsCode deployment, demonstrating measurable improvements in enterprise development efficiency and code quality from production deployment rather than controlled experiments.

\section{Challenges and Lessons Learned}
\label{sec:challenges}

Enterprise AI deployment introduces fundamental challenges that transcend technical implementation. Our 25-month analysis reveals that organizational dynamics, risk management complexity, and systematic model limitations often determine deployment success more than algorithmic performance. 
This section presents empirical evidence contributing to RQ2.

\subsection{Organizational Challenges}
\label{sec:challenges_rq2}
Three organizational challenges emerged consistently across our deployment, providing empirical evidence for RQ2:

\subsubsection{Responsibility Attribution and Risk Asymmetry}
AI-generated code fundamentally disrupts traditional software engineering accountability models. We observed cases where code reviewers assumed no liability and accepted AI diffs as "noop" codemods even when clarified in diff commit summary. 
Code reviewers may also conversely assume full liability for downstream impacts of AI-generated changes. This creates asymmetric risk distribution that alters traditional risk-reward calculations for review processes. 
This dynamic creates systematic disincentives for engaging with complex AI-generated changes, particularly in high-stakes scenarios where human expertise provides maximum value (detailed analysis in Section~\ref{sec:synthesis}).

\begin{table}[t]
\caption{WhatsCode Deployment: Quantitative Impact}
\begin{center}
\scriptsize
\renewcommand{\arraystretch}{1.2}
\begin{tabular}{p{1.6cm}|l|r|r|r}
\hline
\textbf{Category} & \textbf{Metric} & \textbf{Value} & \textbf{Approach} & \textbf{Period} \\
\hline
\multirow{3}{1.5cm}{\textbf{Privacy Compliance}}
& Requirements Mapped & 1,535 & Foundation & 13  \\
& Coverage Improvement & 15\%$\rightarrow$53\% & Foundation & 13  \\
& Automation Diffs & 290 & Foundation & 13  \\
\hline
\multirow{2}{1.5cm}{\textbf{Feature Development}}
& Feature Diffs Assisted & 141 & Agentic & 3 \\ %
& Testing Diffs Assisted & 343 & Agentic & 3 \\ %
\hline
\multirow{4}{1.5cm}{\textbf{Code Quality}}
& Total Diffs Generated & 2{,}000+ & Agentless+ & 12  \\
& Framework Adoption Diffs & 711 & Agentic & 6  \\
& Code Quality Score (Media) & 45\%$\rightarrow$64\% & Agentic & 6  \\
& A/B Expr. Cleanup Diffs & 100+ & Agentic & 6  \\
\hline
\multirow{1}{1.5cm}{\textbf{Adoption}}
& Weekly Active Users & 20\% & Agentic & 6 \\
\hline
\multirow{2}{1.5cm}{\textbf{Operations}}
& Bug Triage Precision & 86\% & Agentless & 6 \\
& Incident Investigation & 21\% & Agentless & 12 \\
\hline
\end{tabular}
\label{tab:metrics}
\end{center}
\vspace{-5mm}
\end{table}

\subsubsection{Algorithmic Aversion and Attribution Bias}
Despite our observations of equivalent bug rates between AI-generated and human-written code, we noticed systematic bias in risk perception for AI-generated failures. Human errors are typically attributed to individual circumstances, while AI errors are interpreted as systematic failures of the entire approach. This attribution asymmetry creates organizational pressure for higher validation standards for AI-generated code, potentially limiting effective utilization of AI capabilities in risk-sensitive contexts.

\subsubsection{Capability Evolution and Expectation Management} 
The rapid advancement of AI capabilities creates fundamental challenges for organizational planning and risk assessment. Tasks seemingly impossible during our foundation era became feasible with agentic infrastructure and stronger base models, requiring continuous recalibration of performance expectations. Organizations should balance conservative deployment strategies with the need to capitalize on emerging capabilities, necessitating agile assessment processes that sample AI performance across diverse problem domains.

\subsection{Limitations and Risk Management}

Enterprise AI deployment reveals systematic failure modes requiring specialized risk management frameworks. Our incident analysis across production deployments identifies four primary failure categories: context misinterpretation, subtle logic errors, unintended side effects, and hallucinated solutions. These failures demonstrate that traditional software validation approaches require enhancement for AI-generated code.

The fundamental challenge lies in detecting semantic correctness beyond syntactic validity. While static analysis effectively identifies structural issues, AI-generated code can pass basic validation while containing logically flawed implementations that manifest only under specific runtime conditions. This necessitates validation protocols including comprehensive testing coverage, provenance tracking for rapid incident attribution, and graduated rollback capabilities specifically designed for AI-generated changes.

A methodological challenge that emerged during our agentless deployment was maintaining systematic evaluation discipline. We observed that rapid iteration to address localized performance failures, while computationally efficient, can introduce systematic regressions that compromise overall system metrics. The low computational overhead of agentless architectures creates opportunities for comprehensive benchmarking protocols that we recommend should be systematically deployed whenever ground truth evaluation datasets are available. This represents an area where more disciplined evaluation methodology could prevent local optimizations from degrading global system effectiveness.

\section{Synthesis and Implications}
\label{sec:synthesis}

This section synthesizes our 25-month empirical study to provide direct answers to our research questions, consolidating insights from across WhatsCode's evolution and deployment experience.

\subsection{Answer to RQ1: Business Impact}

Our systematic measurement across three deployment eras provides evidence of business value achievable through large-scale AI deployment. The results have been presented in each of the Era sections.  
Their quantitative evidence demonstrates that domain-specific AI platforms can achieve substantial, measurable business impact while maintaining organizational acceptance. 
The key insight is that business impact metrics (process automation level, acceptance rates, automated change volume, feature development velocity) provide more meaningful success indicators than traditional productivity metrics alone.

\subsection{Answer to RQ2: Organizational Factors}

Our empirical analysis reveals that organizational factors dominate technical considerations in determining AI tool deployment success. Three organizational dynamics emerged as primary determinants:

\paragraph{Responsibility Attribution Models}
AI-generated code fundamentally disrupts traditional software engineering accountability models. Our analysis reveals that successful deployment requires explicit preservation of human agency within AI workflows. Systems maintaining human involvement throughout the generation process demonstrate superior organizational acceptance (83\% success rate for \powershell fixes with dedicated reviewers vs. 31\% for intent fixes with distributed review) compared to fully autonomous approaches.
The insight is that accountability frameworks must evolve alongside AI capabilities rather than being displaced by them. Organizations should establish clear responsibility attribution where reviewers take ownership of AI-generated code they approve, with explicit escalation and rollback mechanisms for high-risk changes.

\paragraph{Stable Human-AI Collaboration Patterns}
Our longitudinal analysis across 3,000+ accepted code changes reveals two stable collaboration modes that emerged organically from industrial deployment:

\begin{itemize}[leftmargin=*]
\item \textbf{One-click rollout (60\% of interactions):} High-confidence, low-risk scenarios where AI suggestions require minimal human modification. This pattern enables significant productivity acceleration for routine compliance tasks.
\item \textbf{Commandeer-revise (40\% of interactions):} Complex scenarios involving substantial human revision of AI outputs, where developers use AI suggestions as sophisticated starting points for problem-solving.
\end{itemize}

The consistent distribution across diverse tasks and time periods suggests these represent fundamental modes of human-AI collaboration rather than transitional states toward full automation. This finding has implications for system design, suggesting that optimizing for collaborative refinement yields greater long-term value than optimizing for autonomous task completion.

\paragraph{Deployment and Adoption Framework}

Our experience reveals that successful enterprise AI deployment requires coordinated attention to the following three interdependent dimensions:

\noindent \textbf{Architectural Principles:}
Hybrid architectures combining RAG for knowledge grounding, static analysis for validation, and agentic orchestration for complex workflows consistently outperform single-approach systems. The key insight is that enterprise environments require complementary capabilities: RAG provides explainable knowledge access for compliance domains, traditional validation tools ensure correctness, and agent-based systems enable multi-step reasoning for complex tasks. This combination addresses the fundamental mismatch between AI probabilistic outputs and enterprise deterministic requirements.

\noindent \textbf{Organizational Readiness:}
Counter-intuitively, initiating deployment in compliance-relevant domains rather than low-risk applications yields superior adoption outcomes. High-stakes areas often correlate to high frictions and low automation. AI applications provide immediate measurable business value, convert bottleneck owners into efficiency advocates, and establish organizational confidence in AI capabilities. This finding challenges conventional wisdom about gradual technology adoption, suggesting that enterprise AI benefits from demonstrating value in areas where traditional approaches are most constrained.

\noindent \textbf{Risk Management Evolution:}
Traditional software risk assessment requires substantial changes for AI systems. 
We took a graduated autonomy design to carefully calibrate autonomous capabilities based on task risk and organizational readiness. 
Our practice implements risk-aware automation with four levels: autonomous operation for low-risk tasks, supervised automation for medium-risk scenarios, collaborative decision-making for high-risk situations, and human-led assistance for important modifications. 
The insight is that risk assessment must consider both task characteristics and AI confidence, creating dynamic routing mechanisms that optimize human oversight allocation.

\subsection{Takeaways for Enterprise AI Adoption}

The WhatsCode deployment demonstrates that successful enterprise AI adoption depends on alignment between technical capabilities and organizational context rather than raw algorithmic performance. Organizations should simultaneously address technical architecture design, cultural change management, and risk mitigation strategies to achieve sustainable business impact. 
The evidence suggests that human-AI collaboration patterns, rather than full automation, represent the stable equilibrium for enterprise AI systems, requiring system design that optimizes for collaborative effectiveness rather than purely autonomous operations.

\section{Related Work}
\label{sec:related}

\textbf{Surveys and mappings:} The last two years have produced comprehensive overviews of how LLMs are being used across software engineering (SE) tasks. The TOSEM survey by Hou et~al.\ synthesizes 395 papers (2017–2024), cataloging task coverage, datasets, and risks such as evaluation instability, and provides a unifying lens for positioning GenAI-enabled developer tools within the SE lifecycle~\cite{hou2024llm4se}. Fan et~al. maps applications across the SE lifecycle and articulates open problems, emphasizing hybrid approaches that combine traditional SE techniques with LLMs to mitigate hallucinations and improve reliability~\cite{fan2023fose-llmse}.

\noindent \textbf{Coding Assistant and Developer Productivity:}  Microsoft’s multi-study reports and randomized and quasi-experimental evaluations show sizable gains across task speed and collaboration metrics, with heterogeneous effects by role and context \cite{butler2024rct,microsoft2024-gai-workplaces,microsoft2024-nfw}.
Within software product teams, Nahar et al. also surface emerging quality-assurance practices (guardrails, evaluation assets, and LLM-specific workflows) adopted to integrate LLMs into production software \cite{nahar2024-beyond-comfort}.
Khojah et al. documents field observations that practitioners rely on ChatGPT well beyond code generation (e.g., comprehension, refactoring, docs) and adapt their prompting strategies over time~\cite{khojah2024beyond}. 
Industrial adoption studies are also emerging: Davila et al. reports a multi-team case study of AI programming assistants in industry, discussing integration frictions, policy constraints, and perceived productivity effects~\cite{davila2024adoption}.
This growing body of deployment-focused evidence complements earlier lab studies and frames the real organizational constraints that enterprise adopters face.

\noindent \textbf{Testing, debugging and fixing:}
Beyond raw generation, research is reframing developer–LLM interaction around verifiable outcomes. Fakhoury et al. introduce \emph{TiCoder}, a test-driven, interactive workflow that increases pass@1 accuracy and improves user evaluation of AI suggestions~\cite{fakhoury2024ticoder}. On the automation side, Li et al. show that code-aware prompting (SymPrompt) can steer LLMs toward higher-coverage unit tests via coverage feedback~\cite{li2024symprompt}. For post-merge quality, Wadhwa et al present \emph{CORE}, an LLM-based approach to resolve code quality issues mined from production repositories~\cite{wadhwa2024core}. Yang et al. demonstrate test-free fault localization with LLMs, enabling debugging when tests are absent or incomplete~\cite{yang2024llmfl}. Hossain et al. offer a deep empirical analysis of LLMs for bug localization and repair across benchmarks, mapping where prompting and retrieval help versus where models still overfit or hallucinate fixes~\cite{hossain2024deepdive}. Earlier ICSE'23 studies showed both the promise and limits of using (and repairing) LLM-generated code, establishing baselines for today’s enterprise deployments~\cite{xia2023apr,fan2023aprLM}.

While significant work exists on individual components of AI-assisted development, enterprise tool adoption, and compliance automation, there is limited research at their intersection. Our work addresses this gap by providing empirical evidence of large-scale enterprise AI tool deployment with quantitative business impact analysis and comprehensive organizational challenge characterization in compliance-relevant environments.

\section{Threats to Validity}
\label{sec:threats}
Several factors may limit the generalizability of our findings:

\noindent \textbf{External Validity:} Our evaluation is based on a single organization with specific characteristics (messaging platform, privacy-focused, large scale). WhatsApp's specific technology stack, development practices, and regulatory environment may not be representative of other large-scale software projects.

\noindent \textbf{Construct Validity:} Our success metrics focus on compliance automation and productivity gains, which may not capture aspects of software development quality and long-term organizational impact.

\noindent \textbf{Temporal Validity:} AI technology is rapidly evolving, and our findings may become outdated as models improve. However, the organizational and process insights are likely to remain relevant.

\noindent \textbf{Selection Bias:} Our use cases were selected based on clear business needs and technical feasibility, which may overstate AI effectiveness compared to random task selection.
Despite these limitations, we believe this empirical report provides valuable insights for researchers and practitioners considering similar AI tool deployments.

\section{Conclusion}
\label{sec:conclusion}

This paper presents a comprehensive empirical study of enterprise-scale AI development tool deployment, analyzing 25 months of WhatsCode implementation at WhatsApp. Our systematic analysis provides evidence-based answers to fundamental questions about organizational factors determining AI tool deployment success and quantifiable business impact achievable in compliance-relevant industrial environments.

\noindent\textbf{Implications for Practice:}
Organizations should allocate corresponding resources to organizational change management and technical development, with particular attention to ownership attribution frameworks and risk mitigation processes.
Beginning with high developer friction areas that have clear success metrics provides optimal conditions for organizational adoption and measurable business impact demonstration.
Enterprise AI tools should be designed to support both one-click rollout and commandeer-revise patterns, with intelligent routing based on confidence scoring and complexity analysis.
Traditional productivity metrics are insufficient for enterprise AI evaluation; business impact metrics such as compliance coverage, user engagement and feature development velocity provide more meaningful success indicators.

\noindent\textbf{Implications for Research:}
The intersection of AI capabilities and organizational dynamics requires dedicated empirical investigation, moving beyond individual developer productivity studies toward comprehensive organizational impact analysis. Enterprise-specific AI tool evaluation demands new metrics and methodologies that account for compliance requirements, organizational processes, and long-term business impact rather than isolated technical performance measures.
Furthermore, human-AI collaboration patterns in enterprise environments exhibit characteristics distinct from individual developer interactions, requiring specialized theoretical frameworks and empirical investigation methodologies. Future work should focus on developing generalizable frameworks for readiness assessment, standardized metrics for AI impact measurement, and systematic approaches for scaling human-AI collaboration patterns across diverse organizational contexts.

\section*{Acknowledgments}
We thank all WhatsApp engineers for their pilot participation, and the engineering leadership team for strategic guidance and sustained support.
Special thanks to Dino Distefano and Nachi Nagappan for support 
in the kickoff and cross-functional collaborations, and Meta team members (Mark Harman, Jordi Mola, Zhaodong Wang, Nathan Hu) for technical and collaborative support.

\bibliographystyle{ACM-Reference-Format}
\bibliography{citations.bib}

\end{document}